# Music Between Two Minds: Individual Signatures, Shared States, and the Emergence of Creative Interaction

Tatsuya Daikoku[1,2,*], Risa Kobayashi[1,3], Rina Tabuchi[1], Yoshia Abe[1], Yasuo Kuniyoshi[1]

1 Graduate School of Information Science and Technology, The University of Tokyo, Tokyo, Japan

2 Centre for Neuroscience in Education, University of Cambridge, Cambridge, UK

3 Faculty of Child Studies, Seitoku University, Japan

*** Corresponding author**

Tatsuya Daikoku

Graduate School of Information Science and Technology, The University of Tokyo, 7-3-1 Hongo, Bunkyo-ku, 113-8656, Tokyo, Japan

Email: daikoku.tatsuya@mail.u-tokyo.ac.jp

# Abstract

Some of the most valued forms of human creativity arise between people—in conversation, collaboration, and improvised music—yet how two minds jointly regulate the flow of novelty in real time remains unknown. Here we examined the information dynamics of interpersonal improvisation using 986 professional jazz improvisations. Pitch-based fluctuations of uncertainty and surprise behaved as signatures of individual identity: the two partners differed within 90% of takes, performers were identifiable from single solos and the separation between partners persisted undiminished across turns and takes. Rhythm-based fluctuations showed the reverse organization: partners were statistically indistinguishable within a take yet shifted together from take to take, and this shared state drifted gradually as each performance progressed, with each performer moving away from their own past at the same rate as from their partner's. Beyond joint drift, solo hand-offs revealed turn-by-turn complementarity: when one performer exceeded their own average surprise in the predictive dynamics of improvisation, the partner's next solo fell below their own, in both pitch and rhythm and in all duos. Improvising pairs thus neither fuse into one system nor remain two independent ones: they preserve individual melodic voices while sharing a rhythmic state, within which moment-to-moment push–pull continually rebalances their interaction. Individuality and togetherness thus coexist, carried by different musical dimensions—a measurable organization of creativity that emerges between minds.

*Keywords:* creativity; social interaction; predictive processing; improvisation; collective intelligence; field intelligence

## Significance Statement

Creativity is usually attributed to individuals, yet much of what people create—music, conversation, discoveries—is improvised together. Using a computational model of musical expectation as a common yardstick, we analyzed how the predictability of every note fluctuates between musicians trading improvised solos. Two forms of organization coexisted in the duos. Melodic fluctuations identified who was playing and did not converge. Rhythmic fluctuations instead belonged to the performance itself: shared by both partners, drifting slowly as the take progressed, and rebalanced at every exchange through a push–pull in which, when one player's surprise was higher than usual, the partner's next solo tended to be less surprising. Creative interaction thus preserves individual identity while co-creating a shared state, offering a quantitative window onto how novelty is regulated between minds.

# Introduction

From improvised conversation to scientific collaboration to music, some of the most valued human creativity arises between people, in real time. Creativity is commonly defined as the capacity to generate something both novel and valuable (Hennessey and Amabile, 2010; Runco and Jaeger, 2012). Yet novelty alone cannot explain why some deviations are recognized as creative whereas others are dismissed as noise. Creative deviations may instead be those that are sufficiently unexpected to be informative, yet sufficiently interpretable within the predictive frameworks of individuals and cultures (Huron, 2006; Clark, 2013). Consistent with this view, creativity emerges not only from isolated individuals but also from collective processes through which ideas are selected, refined, and transmitted across groups and social networks (Uzzi and Spiro, 2005; Wuchty et al., 2007; Woolley et al., 2010; Muthukrishna and Henrich, 2016). What such accounts leave unresolved, however, is how creative coordination is achieved in the moment—when two minds must generate novelty together, turn by turn.

Predictive processing offers a principled framework for addressing this question. The brain continuously generates predictions about incoming events from prior experience and updates them in response to sensory input (Friston, 2010; Clark, 2013). One biologically grounded capacity supporting these processes is statistical learning: the extraction of regularities from the environment through exposure (Saffran et al., 1996). In music, listeners implicitly acquire the statistical regularities of pitch, rhythm, and harmony, and use them to anticipate upcoming events (Pearce and Wiggins, 2012; Koelsch et al., 2019; Daikoku, 2018). Because music unfolds through time, musical experience depends not simply on isolated moments of prediction error, but on the evolving temporal dynamics of uncertainty and surprise—how expectations are established, violated, and resolved within an ongoing context (Salimpoor et al., 2011; Cheung et al., 2019). Crucially, these predictive quantities can be computed for every event in a symbolic sequence, making interaction measurable at the resolution of single notes.

Improvised jazz provides an ideal arena for such measurement. Improvisers draw on learned musical regularities while introducing subtle prediction violations in real time, and the practice of trading—two performers alternating improvised solos over a shared form—imposes a conversational turn-taking structure on the exchange. A recent corpus analysis of 456 jazz improvisations by 78 musicians from 1925 to 2009, using a statistical-learning model, showed that although acoustic features changed little across historical periods, the temporal dynamics of surprise and uncertainty in pitch sequence exhibited clear period-specific patterns (Daikoku, 2024). The history of jazz can thus be read as the collective evolution of shared predictive fluctuations that musicians gradually transformed over time—raising the question of how such fluctuations are generated and regulated between individuals in the first place.

How co-actors coordinate during live creative interaction, however, has mainly been studied outside the framework of predictive processing. Research on dialogue shows that interacting people align their linguistic and conceptual representations (Pickering and Garrod, 2004) and couple their neural dynamics during communication (Hasson et al., 2012). Studies of joint musical improvisation have characterized coordination in body movement—spontaneous synchronization and shared periodic motion between co-performers (Walton et al., 2015; Eerola et al., 2018)—and in the communicative feedback that improvising partners exchange (Moran et al., 2015). Yet creative interaction poses a puzzle that alignment alone cannot solve: partners who fully converge produce nothing new, whereas partners who ignore each other produce no interaction at all. Whether and how co-performers coordinate in the informational space of prediction—the moment-to-moment surprise and uncertainty that their choices generate under shared cultural knowledge—has not, to our knowledge, been directly quantified at the level of note-by-note predictive uncertainty and surprise.

In a recent theoretical framework, we proposed that creative interaction is sustained by field intelligence: the generation, regulation, and selection of predictive fluctuations that cannot be attributed to either partner "alone", but become intelligible only at the level of their "interaction" (Daikoku, 2026). On this view, the critical problem is not simply whether interacting individuals align, but how a pair can preserve the diversity necessary for novelty while preventing their deviations from becoming mutually unintelligible. Field intelligence therefore predicts an organization in which individual differences are not erased by interaction, yet some aspects of predictive dynamics become jointly organized through the interaction itself. Such organization should be detectable not merely as overall similarity between partners, but as structure that is specific to the particular occasion of playing together, together with moment-to-moment dependencies that cannot be reconstructed from the two performers considered independently. This reframes collective creativity as a question of where predictive fluctuations reside—within individuals, within the interaction, or in the ongoing relation between the two—and how these levels coexist and change over time.

The present study provides a first empirical test of this proposal using a corpus of 986 Jazz improvisations. We trained a general predictive model of jazz expectation on the Jazzomat corpus of 456 professionally transcribed solos (Frieler et al., 2013) and then held it fixed while applying it, note by note, to the Improvising Duos corpus (Moran et al., 2015)—specifically, to the solo-trading sections of pulsed performances of the jazz standard *Autumn Leaves*, in which two performers strictly alternate improvised solos (6 duos, 12 professional improvisers, 54 takes, 530 solos, 11,433 notes). For every note, the model yields the surprise of the realized event and the uncertainty of its context, separately for pitch, rhythm, and their conjunction. We asked three questions: (i) are interpersonal predictive fluctuations organized by stable individual signatures (who is playing) or by emergent performance-level states (which performance is unfolding); (ii) how do these fluctuations evolve as a performance unfolds; and (iii) is there turn-level coupling between partners beyond what two independent performers would produce?

We hypothesized that in expert duos the partners' predictive fluctuations would neither fully coincide nor drift apart independently (Daikoku, 2026). Rather, the field-intelligence account predicts the coexistence of preserved individuality and interaction-level organization: predictive dynamics may retain signatures of the individuals who generate them while, at the same time, acquiring structure that is specific to the unfolding interaction. The critical question is therefore not whether two improvisers become similar or remain different, but how individual and shared predictive organization coexist and evolve as they create novel information together. By distinguishing performer-specific, performance-specific, and turn-dependent components of predictive dynamics, the present study tests whether creative interaction gives rise to an organization that cannot be reduced to either performer considered alone.

## Results

### Quantifying predictive fluctuations between improvising minds

To measure the predictive fluctuations that improvisers generate for one another, we first constructed a general model of jazz expectation and then let it listen to real duo interaction, continuing to update its statistics as a listener accumulating experience would. A hierarchical

Bayesian statistical-learning model (HBSL; Daikoku et al., 2023) was trained on the Weimar Jazz Database of the Jazzomat project—456 professionally transcribed solos by 78 musicians (Frieler et al., 2013)—separately for three event streams: pitch, rhythm, and their conjunction (Materials and Methods). For every note of a new performance, the model outputs the surprise of the realized event (information content: the negative log probability of the note given its context) and the uncertainty of that context (the entropy of the model's predictive distribution), both in bits. Because the model's knowledge came primarily from the Jazzomat corpus and was updated only gradually by the duo recordings, these quantities provide a common, corpus-level yardstick of idiomatic jazz expectation. The two partners within a take were evaluated by the model at essentially the same state, so comparisons between them are not affected by this gradual updating, and slower session-wide change was controlled for by comparisons with the immediately adjacent take (see Materials and Methods). We applied this fixed model to the Improvising Duos corpus (Moran et al., 2015), focusing on the solo-trading sections of pulsed performances of *Autumn Leaves*, in which the two performers strictly alternated improvised solos. After screening (Materials and Methods), the dataset comprised 6 duos (12 professional improvisers; one duo each of bass–guitar, piano–bass, piano–saxophone, piano–violin, saxophone–electric bass, and trumpet–guitar), 54 takes, 530 solos, and 11,433 note events (10,903 with a defined predictive context).

**Melodic fluctuations index who is playing**

Pitch-based predictive fluctuations behaved as stable signatures of individual identity (Fig. 1). Within a single take, the note-level uncertainty values of the two partners differed significantly in 90% of takes, whereas the same performer compared across consecutive takes differed in only 35%; this contrast held in all six duos ($t(5) = 6.40$, $P = 0.0014$). The separation between partners was large and persistent. That is, within a take, distances between solos by the two different performers were 2.14 times greater than distances between solos by the same performer (mean Jensen–Shannon divergence 0.261 versus 0.125; $t(5) = 2.87$, $P = 0.035$; 6/6 duos), and the distance to the partner's solos was as large one turn later as five turns later (0.270 versus 0.254; $P = 0.110$; Fig. 1a, top; Fig. 1b, left)—a gap that turn-taking never closed. Performer identity could be decoded from single solos (leave-one-take-out cross-validation; AUC = 0.87; above chance in 6/6 duos; $t(5) = 7.78$, $P = 0.0006$), and PERMANOVA on solo-to-solo distance matrices attributed significant variance to performer in every duo ($R^2 = 0.13$–0.81) but to take in only one ($R^2 = 0.01$–0.33), with performer exceeding take in 5/6 duos (mean $R^2 = 0.44$ versus 0.11; paired difference $P = 0.088$). Converging on the same conclusion, a mixed-model variance decomposition of solo means assigned 74.3% of the variance in uncertainty to performer identity and only 0.6% to take (surprise: 18.6% versus 1.3%). In melodic space, in short, the corpus knows who is playing. Because the two members of each duo played different instruments, part of this signature may reflect instrument-specific affordances for pitch; we return to this point in the Discussion.

**Rhythmic fluctuations index which performance is unfolding**

Rhythmic predictive fluctuations showed the opposite organization. Within a take, the two partners were largely indistinguishable: only 39% of take-level comparisons reached significance, despite the same note-level statistical power that detected pitch differences in 90% of takes. Instead, rhythmic fluctuations changed from performance to performance: the same performer differed from themselves across consecutive takes in 60% of comparisons, significantly more often than the two partners differed within a take ($t(5) = 2.74$, $P = 0.041$; 5/6 duos). Between-performer and within-performer distances were equivalent (ratio 0.99; 0.136 versus 0.139; $P = 0.62$), performer identity could not be decoded from rhythmic fluctuations (AUC = 0.52, $P = 0.72$), and PERMANOVA attributed significant variance to take in every duo ($R^2$ = 0.27–0.85) but to performer in none ($R^2$ = −0.01–0.03; significant in 0/6 duos), with take exceeding performer in 6/6 duos (mean $R^2$ = 0.56 versus 0.01; paired difference $P$ = 0.0018).

Critically, the take structured both performers together, and only performers who actually played together shared it. We compared each solo's distance to the partner's next solo in the same take (real interaction) with its distance to the same partner instrument's solo at the same turn position in a different take (a pseudo-duet of the same two people who were not, at that moment, playing together). Rhythmic distances were twice as large in the pseudo-duets (0.116 versus 0.238; $t(5)$ = −3.68, $P$ = 0.014; 6/6 duos), and the difference persisted when pseudo-duets were drawn only from the immediately adjacent take (0.116 versus 0.161; $t(5)$ = −4.04, $P$ = 0.0099; 6/6 duos), ruling out slow session-wide change as the sole explanation. Pitch showed no such take-specificity (0.275 versus 0.280; $P$ = 0.31). The joint pitch–rhythm stream patterned with pitch throughout (e.g., within-take versus across-take significance rates 63% versus 26%, $P$ = 0.034; decoding AUC = 0.78, $P$ = 0.0067), consistent with melodic information dominating the conjoint representation. Whereas melodic fluctuations answer who is playing, rhythmic fluctuations answer which performance is unfolding: each take establishes a rhythmic state that both performers jointly occupy—a property of the pair and the occasion that cannot be assigned to either individual.

**The shared rhythmic state drifts gradually and collectively**

If each take establishes a shared rhythmic state, does that state stay put or move? We tracked how the dissimilarity between two solos grew with the number of turns separating them, exploiting the strict alternation of trading: odd separations (1, 3, 5 turns apart) always compare a performer with their partner, whereas even separations (2, 4, 6 turns apart) always compare a performer with themselves, so the two series can be read directly as distance to the other and distance to one's own past (Fig. 1). In rhythm, both series climbed together. The distance to the partner rose from 0.111 (1 turn apart) to 0.126 (3 turns) to 0.146 (5 turns) ($t(5) = 2.54$, $P = 0.052$; 5/6 duos), and the distance to one's own previous solos rose in parallel from 0.116 to 0.134 to 0.149 ($t(5) = 4.65$, $P = 0.0056$; 6/6 duos); across all available separations (up to 22 turns), the lag–distance correlation was positive in every duo (mean Spearman $\rho = +0.17$; $t(5) = 3.89$, $P = 0.012$; 6/6 duos) (Fig. 1a, bottom; Fig. 1b, right). Because each performer moved away from their own past at the same rate as from their partner, the pair was not diverging: the two performers were travelling together, and the shared rhythmic state itself drifted gradually across the take. Pitch showed no such drift (mean Spearman $\rho = +0.01$, $P = 0.76$; increase in partner distance, $P = 0.110$, 1/6 duos): the melodic separation between partners was as constant in time as it was wide (Fig. 1b, left). Gradual joint drift alone, however, does not demonstrate interaction—two performers independently wandering at similar rates could produce a similar profile. Evidence for genuine interpersonal coupling requires turn-level dependence between the partners, which we tested next.

**Solo hand-offs reveal turn-by-turn complementarity**

Solo hand-offs revealed a systematic push–pull between partners (Fig. 2). For each solo we computed its mean surprise, removed each take's linear trend, and expressed the residual relative to that performer's own average, asking whether a performer's deviation on one turn predicted the partner's deviation on the next (Materials and Methods). Because strict alternation combined with stable level differences between performers can produce spurious negative correlations, we benchmarked the observed coupling against two null models that preserve everything about each individual's playing except its alignment with the partner: shuffling each performer's solos within their own turn positions (control A) and circularly shifting one performer's sequence against the other's (control B)—a world in which the same two people played the same music, but out of step. Real duos showed reliable negative coupling that both controls abolished. In rhythm, the mean within-duo correlation across consecutive solos was $r = -0.230$, versus $-0.000$ under control A and $+0.069$ under control B ($t(5) = -11.24$ and $-11.50$; both $P < 0.0001$; 6/6 duos); in pitch, $r = -0.151$ versus $+0.001$ and $+0.045$ ($t(5) = -4.33$ and $-4.47$; $P = 0.0075$ and $0.0066$; 6/6 duos); in the joint stream the effect was weaker and did not reach significance ($r = -0.110$; $P = 0.105$ and $0.084$; 5/6 duos). Pooling all hand-offs in the corpus ($n = 415$), a performer who played above their own average was reliably followed by a partner playing below their own (rhythm, $r = -0.19$; pitch, $r = -0.10$; Fig. 2).

A complementary, unsigned analysis asked whether the partner's temporally adjacent solo was also the most similar. This held for rhythm only: solos were closer to the partner's next solo than to the partner's solos three or more turns away (Jensen–Shannon divergence, 0.114 versus 0.147, $t(5) = -3.09$, $P = 0.027$, 5/6 duos; Kolmogorov–Smirnov statistic, $t(5) = -4.54$, $P = 0.0062$, 6/6 duos), but not for pitch ($P = 0.116$ and $P = 0.253$, respectively). The two analyses are not redundant: the signed push–pull captures the direction of momentary deviations, whereas the unsigned analysis captures overall proximity, and a deviation answered by a counter-deviation increases directional coupling while leaving the partners' distributions apart. Together they indicate a shared rhythmic state that is continuously re-negotiated at each exchange, alongside melodic identities that remain apart even while the performers push and pull against each other's momentary deviations.

In sum, five independent signatures converged on a single organization. Melodic fluctuations carried a persistent individual identity: partners never converged, their separation neither narrowed with turns nor shifted with takes, and yet they responded to each other's momentary deviations. Rhythmic fluctuations carried a collective state: indistinguishable between partners, specific to each real interaction, drifting gradually within each performance, and rebalanced turn by turn at every hand-off.

## Discussion

We examined how the predictive fluctuations of two improvising musicians organize between them. The results revealed four main patterns. First, pitch-based fluctuations of uncertainty and surprise were signatures of individual identity: the two partners differed in nearly every take, single solos sufficed to identify who was playing, and the separation between partners persisted undiminished across turns and takes. Second, rhythm-based fluctuations were properties of the performance rather than individual identity: partners were indistinguishable within a take, shifted together between takes, and were markedly closer to the partner they were actually playing with than to the same partner in a pseudo-duet assembled from another take. Third, this shared rhythmic state was not static but drifted gradually across each performance, with both performers moving together. Fourth, at every solo hand-off the partners showed a push–pull complementarity—in both pitch and rhythm—that vanished under null models preserving each individual's playing while breaking its alignment with the partner. Together, these results provide a first quantitative characterization of how the flow of predictability is regulated between minds during naturalistic creative interaction.

The central pattern is a double dissociation that resists both classical pictures of joint performance. If the two performers had merged into a single system, they should have been as indistinguishable in pitch as they were in rhythm; if they had remained two independent systems, rhythm should have tracked the individual as reliably as pitch did. Instead, every duo realized both organizations at once, in different dimensions. Put simply, each performer held on to their own way of speaking—their melodic voice—until the very end, while sharing the rhythmic feel of the moment completely with their partner; and on top of that shared state, a push–pull continued at every exchange. Identity and communion were not competing outcomes of the interaction but simultaneous achievements, carried by different musical dimensions.

The shared rhythmic state offers a measurable instance of what we have theoretically termed field intelligence (Daikoku, 2026): structure in predictive fluctuations that cannot be attributed to either partner alone. Variance decomposition made this concrete. The variability of rhythmic fluctuations was essentially unexplained by who was playing and strongly explained by which performance was unfolding—and the pseudo-duet analysis showed that this state belonged to the actual interaction, not to the pair as a fixed combination of two people. What carries the state is therefore neither performer, but the particular occasion of playing together: a property that exists only in the field between them.

The hand-off analysis further shows that this shared organization is actively regulated rather than passively inherited. A performer who exceeded their own average surprise was answered by a partner falling below their own. This is complementarity, not imitation—the opposite of the convergence emphasized in accounts of interpersonal synchrony and alignment. One interpretation is that the pair co-regulates the joint flow of unpredictability: when one spends, the other saves, keeping the total stream of surprise within a workable band while the shared state slowly explores. On this reading, the drift of the rhythmic state and the turn-level push–pull may be two faces of the same process: collective movement made safe by mutual compensation. These findings also refine our original hypothesis. We had proposed that creative duos maintain an interpersonal distance in predictive fluctuations that is neither too close nor too far (Daikoku, 2026). The results show that this intermediate organization is real but is not implemented as a compromise on a single dimension. It is implemented as a division of roles across dimensions: complete sharing in one (rhythm), preserved difference in the other (pitch), and active rebalancing at the point of exchange in both. A duo can thereby be maximally close and reliably distinct at the same time—without either dimension settling at an intermediate value. Projected onto a single axis, such an organization would resemble an inverted-U relation (Timmons et al., 2015) between interpersonal distance and interaction quality, but the present study did not test that relation directly; establishing it will require simulations that compare candidate interaction models and perceptual experiments that manipulate interpersonal fluctuation distance in synthesized performances. For the same reason, our data show how novelty is jointly regulated, but they do not yet show that this regulation is what makes the interaction creative in the evaluative sense; linking the present dynamics to judgments of creativity and aesthetic value is a necessary next step.

A further principle suggested by the data is that what is shared moves, and what is not shared is preserved. The dimension in which the partners fully converged—rhythm—was precisely the dimension that travelled across the take and reset between takes, whereas the dimension that carried stable identity—pitch—stayed where it was. This division may be functionally important for collective creativity in general. A group that moves together in some shared dimension can explore new territory without fragmenting, while stable individual differences maintain the diversity of material available for exploration. In this light, the improvising duo realizes in miniature what theories of collective intelligence and cultural evolution describe at larger scales: innovation arising from the interplay of shared states and preserved individual variation (Uzzi and Spiro, 2005; Woolley et al., 2010; Muthukrishna and Henrich, 2016).

Although our data are musical, the organization they reveal is stated entirely in domain-general terms—identity-bearing dimensions, occasion-specific shared states, collective drift, and complementary rebalancing at turn exchanges—and each has a natural analogue in conversation and other turn-taking interactions. Dialogue partners align many aspects of their language (Pickering and Garrod, 2004) and couple neurally during communication (Hasson et al., 2012), yet speakers plainly retain personal voice and style across conversations, and every conversation has a feel of its own that both parties inhabit and jointly steer. Our results predict that, given a suitable corpus-trained reference model, the same dissociation should be measurable in dialogue: some informational dimensions of speech should identify the speaker and resist convergence, others should belong to the particular conversation and drift with it, and turn exchanges should show compensatory regulation of surprise. Music offered a privileged test bed for this program—symbolic, culturally codified, and strictly alternating—but the analytic framework transfers to any interaction that can be represented as a sequence of events under a shared predictive model.

Several limitations should be acknowledged. First, the two members of every duo in this corpus played different instruments, so the melodic identity signature may partly reflect instrument-specific affordances rather than personal style alone: a guitarist may remain guitar-like across takes for reasons of register and idiom as much as of individuality. Dissociating person from instrument will require corpora with same-instrument duos or performers recorded on multiple instruments. Notably, this concern does not touch the rhythmic findings, whose defining feature is precisely the absence of individual (and hence instrument) specificity. Second, the analyses are observational: they establish that the observed coupling exceeds stringent null models, but not the causal pathway that produces it. Third, all performances were improvisations on a single tune within a single idiom; whether the same division of roles between identity-bearing and state-bearing dimensions holds across repertoires, genres, and cultures—and whether the assignment of pitch to identity and rhythm to the shared state is universal or idiom-specific—remains to be tested. Fourth, the reference model was trained on a specific historical corpus of monophonic solos; different reference corpora could shift absolute values of surprise and uncertainty, although the within-corpus contrasts on which all conclusions rest should be robust to this choice. These limitations are balanced by a key strength: to our knowledge, this is the first corpus-wide analysis of note-level predictive dynamics in the Improvising Duos dataset, and all conclusions rest on duo-level inference with effects consistent in five or six of six duos across independent analysis families.

Together, these findings recast the question of whether improvising partners become one system or remain two. The answer depends on the musical dimension considered: individual identity is preserved in one dimension, whereas a shared and dynamically regulated state emerges in the other dimension, with both dimensions further shaped by turn-by-turn rebalancing. Interpersonal improvisation therefore appears to depend not on convergence toward a single common state, but on the simultaneous maintenance of individual and shared predictive dynamics. This organization provides a measurable account of how creative interaction can preserve individual identity while generating shared and novel structure that emerges only at the level of the dyad.

# Materials and Methods

### Corpora and performance selection

Two corpora (986 samples of Jazz improvisation) served complementary purposes. To construct a general predictive model of jazz improvisation, we used the Weimar Jazz Database of the Jazzomat Research Project (https://jazzomat.hfm-weimar.de), a large-scale symbolic database of 456 professionally transcribed monophonic jazz solos by 78 musicians (Frieler et al., 2013). This corpus was used exclusively to estimate population-level statistical regularities of the jazz idiom and was never fitted to the performances used for hypothesis testing.

To examine interpersonal predictive dynamics during real-time interaction, we analyzed the Improvising Duos corpus (Moran et al., 2015; https://doi.org/10.1371/journal.pone.0130070) performing both pulsed improvisations based on the jazz standard *Autumn Leaves* and non-pulsed free improvisations. We focused on the pulsed *Autumn Leaves* performances because they provide a common musical framework across duos and can be converted into reliable symbolic representations; the non-pulsed free improvisations were excluded because the absence of a stable metrical structure limits the reliability of symbolic transcription. Recordings in which the piece or performance style actually played deviated from what the experimenters had instructed were excluded. Within the retained recordings, we analyzed the solo-trading sections, in which the two performers strictly alternated improvised solos over the shared form. The boundaries of each improvisational episode were identified and annotated by three trained musician-authors (R.K., R.T., and T.D.), and each trading section was segmented into successive solos. The final dataset comprised 6 duos (12 performers; one duo each of bass–guitar, piano–bass, piano–saxophone, piano–violin, saxophone–electric bass, and trumpet–guitar), 54 takes, 530 solos, and 11,433 notes. Because individual analyses impose additional structural requirements (for example, a minimum number of notes per solo or a minimum number of solos per take; see Statistical analysis), the exact numbers entering the analyses are given in Statistical analysis.

### Audio source separation and symbolic transcription

All recordings were processed with a uniform pipeline to obtain performer-specific symbolic note sequences. Audio was amplitude-normalized, transformed into short-time Fourier representations, and separated into two sources using AuxIVA and ILRMA as implemented in *pyroomacoustics* (Ono, 2011; Kitamura et al., 2016; Scheibler et al., 2018), with the resulting signals assigned to the two performers by auditory inspection, spectrogram inspection, and instrument identity. Because blind separation did not consistently yield sufficiently clean performer-specific signals, the final analyses used LALAL.AI (https://www.lalal.ai/), an AI-based stem-separation system, to extract performer-dominant audio; this choice is consistent with the broader methodological shift toward deep neural source separation in music information retrieval (Hennequin et al., 2020; Défossez, 2021; Rouard et al., 2023). When direct-line recordings were available, these were used as performer-specific references. The goal of this step was not to benchmark separation algorithms or obtain perfectly isolated stems, but to derive performer-dominant signals sufficient for transcription; separated signals were validated by auditory and spectrogram inspection.

Separated waveforms were converted into MusicXML using Klangio Transcription Studio (https://klang.io/), a browser-based AI-assisted transcription platform, and the resulting scores were manually inspected and edited in notation software (MuseScore 2), with meter, tempo, note durations, and rhythmic quantization adjusted to obtain clean symbolic representations. Because the analyses concern symbolic note sequences rather than expressive microtiming, minor rhythmic deviations present in the raw audio were regularized at this stage. All solo parts were reduced to monophonic melodies. The MusicXML files were temporally aligned with the corresponding audio using dynamic time warping, and phrase-level pitch estimation was refined with a Viterbi dynamic-programming algorithm integrating (i) an acoustic likelihood derived from the fundamental frequency, harmonic structure, local spectral contrast, and the explainability of candidate pitches by lower fundamentals, (ii) melodic transition probabilities, (iii) bidirectional melodic context from neighboring notes, and (iv) a weak prior derived from the original MusicXML. Candidate pitches were evaluated over complete phrases rather than note by note, suppressing octave errors, harmonic overtones, and transient resonance peaks; where simultaneous notes occurred in solo parts, only the pitch most strongly supported by the audio and the surrounding melodic context was retained.

**General predictive model and note-level measures**

Each symbolic solo was parsed into three complementary event streams: a pitch stream (the ordered succession of pitch events), a rhythm stream (the ordered succession of duration and onset-related events), and a joint stream encoding each tone as a compound pitch–rhythm symbol (Daikoku, 2024). To establish a common tonal reference, all pieces—in both the training and the interaction corpus—were transposed to C major or A minor. Predictive quantities were computed with a hierarchical Bayesian statistical-learning (HBSL) model (Daikoku et al., 2023), following previous applications to musical expectation and jazz improvisation (Daikoku, 2024). The model estimates the conditional probability of an upcoming event given its preceding context; each transition is encoded as a beta-distributed belief that is recursively updated as observations accumulate, and predictive probabilities correspond to posterior predictive probabilities derived from these beliefs. Although the model can recursively group recurrent transitions into higher-order chunks, this mechanism was disabled: all analyses were based exclusively on note-to-note transition statistics, so as to quantify local predictive dynamics. The model was first trained on the Jazzomat corpus and then continued to update its transition statistics online (memory-keeping mode) while evaluating the duo corpus, as a listener accumulating experience would. Because this makes the model's state change slowly over the course of the session, take-specific effects were additionally tested against pseudo-duets drawn from the immediately adjacent take, which cancels slow session-wide change (see Statistical analysis).

For each event, surprise was indexed by information content,

$$I(e_{t+1}) = -\log_2 P(e_{t+1} \mid e_t),$$

where $e_{t+1}$ denotes the observed event and $e_t$ its preceding context, and uncertainty was indexed by the entropy of the model's predictive distribution,

$$H(e_{t+1}) = -\sum_{e \in S} P(e_{t+1} = e \mid e_t) \log_2 P(e_{t+1} = e \mid e_t),$$

where $S$ denotes the set of possible events. High information content indicates that the realized event was unlikely under the learned model of the jazz idiom, whereas high entropy indicates that the context afforded many plausible continuations. Both quantities are expressed in bits and were computed for every note of every solo, separately for the pitch, rhythm, and joint streams.

## Statistical analysis

*Inference framework.* The duo was the unit of statistical inference ($n$ = 6). Analyses were restricted to the solo-trading sections of takes containing at least six trading solos (46 takes, 461 solos, 7,225 notes). Effects were first computed within each duo and then tested across duos with two-tailed paired (or one-sample) $t$-tests (df = 5). Given the small number of duos, alongside each test we report the number and ratio of duos showing the effect in the same direction, which provides a distribution-free index of consistency that does not rely on the normality assumption of the t-test. Because note-level distributions of model-derived quantities are strongly skewed, comparisons of raw note-level values used the Mann–Whitney *U* test. All analyses were performed separately for the pitch, rhythm, and joint streams.

*Note-level dissociation of performer and performance (significance rates).* For each take, the note-level uncertainty values of performer A (all notes of all their solos in that take) were compared with those of performer B in the same take (within-take, between-performer family; $n = 45$ comparisons), and each performer's values in a given take were compared with the same performer's values in the following take (between-take, within-performer family; $n = 78$ comparisons); comparisons in which either set contained fewer than 15 notes were skipped. For each duo we computed the proportion of significant comparisons ($\alpha = 0.05$) in each family and compared the two proportions across duos with a paired *t*-test. Because every note enters the test directly (on average ~16 notes per solo), no information is lost to averaging; take-level means shown in figures are for display only.

*Solo-level distances.* For distribution-based analyses, each solo containing at least five notes was represented by the distribution of its note-level values, discretized into eight quantile bins defined on all trading solos of the corpus (with 0.5 added to every bin), and dissimilarity between solos was quantified with the Jensen–Shannon divergence (JSD; Lin, 1991). JSD was chosen because it compares entire distributions rather than their means, is symmetric and bounded (0–1 bits), and remains finite when a bin is empty in one of the two solos. Within each take, mean JSD between solos by the two different performers and mean JSD between solos by the same performer were computed per duo, and the two means were compared with a paired t-test across duos, and their ratio (between/within, averaged across duos) is reported as an effect size.

*Performer decoding.* A leave-one-take-out cross-validated classifier predicted, from a single solo, which of the two duo members had played it. Features were five summary statistics (mean, s.d., median, maximum, minimum) of each of the two note-level measures (surprise and uncertainty; ten features in total); solos for which the s.d. was undefined (a single note) were excluded. For each held-out take, an L2-regularized logistic regression (with feature standardization; C = 0.5) was trained on all remaining takes of that duo, and out-of-fold predicted probabilities were pooled to compute one area under the ROC curve (AUC) per duo. Duo-level AUCs were tested against chance (0.5) with a one-sample *t*-test.

*PERMANOVA and variance components.* Within each duo, the solo-to-solo JSD matrix (solos with at least five notes; duos with at least 20 such solos) was Gower-centered, and the variance explained by performer (adjusted for take) and by take (adjusted for performer) was computed via projection matrices, yielding partial $R^2$ values for the two factors. Significance was assessed with restricted permutations (999 iterations): performer labels were permuted within takes for the performer test, and take labels were permuted within performers for the take test, with $P$ = (number of null statistics ≥ observed + 1)/(999 + 1). The duo-level difference between the two partial $R^2$ values was tested with a paired *t*-test. Complementarily, a linear mixed model decomposed the variance of solo-level means (standardized within each measure): fixed effects of take order, turn order within the take, and *z*-scored note count; a random intercept for duo; and variance components for performer (duo × instrument) and take, estimated by restricted maximum likelihood. Reported percentages are each component's share of the total variance.

*Lag-resolved drift.* For every pair of solos within a take, the lag was defined as the number of turns separating them. Because trading strictly alternates, odd lags (1, 3, 5) always compare a performer with their partner and even lags (2, 4, 6) always compare a performer with themselves; the two strata were therefore never mixed. For the lag-binned analysis, mean JSD at each lag was computed per duo and stratum, and the change across lags was tested by comparing the largest with the smallest lag (lag 5 versus 1 for the partner stratum; lag 6 versus 2 for the self stratum) with paired *t*-tests. Because takes differ in length (6–23 solos), the figure and the lag-binned statistics were restricted to takes able to supply every lag from 1 to 6 (41 of 46 takes with analyzable solo pairs), so that identical takes underlie every lag; conclusions were unchanged without this restriction. As a trend analysis using all available lags (up to 22), Spearman correlations between lag and distance were computed within each take and stratum (requiring at least six pairs and three distinct lags), averaged per duo, and tested against zero with a one-sample *t*-test.

*Take-specificity of the shared state (pseudo-duets).* For each hand-off within a take, the distance between a solo and the partner's next solo (REAL) was compared with the mean distance between the same solo and the solos of the same partner instrument at the same turn position in the other takes (FAKE)—a pseudo-duet pairing of performances that did not occur together. Duo-level means of REAL and FAKE were compared with paired *t*-tests, both across all takes and in a control restricted to the immediately adjacent take, which cancels the influence of slow session-wide change and of the temporal separation itself.

*Hand-off coupling and null models.* Each solo was summarized by its mean surprise. Within each take with at least six solos and exactly two performers, the sequence of solo means was linearly detrended, and each performer's mean was subtracted from their own solos, yielding a sequence of deviations relative to each performer's own average at that point in the take. The push–pull statistic was the Pearson correlation between consecutive deviations (which always belong to different performers), averaged per duo. Two null models preserved each individual's playing while destroying its alignment with the partner: control A permuted each performer's deviations within that performer's own turn positions (2,000 random permutations, averaged); control B circularly shifted one performer's sequence by every possible offset (averaged over offsets), preserving each performer's own temporal structure exactly. Real and null coupling values were compared across duos with paired *t*-tests. This test is sensitive to the construction of the null hypothesis: without subtracting performer means, stable level differences between the two performers dominate the covariance and mask turn-level coupling (producing a false negative for pitch); performer-mean centering removes this artifact. For visualization and a pooled estimate, all hand-offs in the corpus ($n$ = 415) were plotted as (outgoing performer's deviation, incoming performer's deviation) pairs and summarized with a Pearson correlation.

*Unsigned adjacency.* To test whether the partner's temporally adjacent solo was also the most similar irrespective of the direction of deviations, distances between each solo and the partner's solos at lag 1 were compared with distances to the partner's solos at lags $\geq 3$, using both JSD and the two-sample Kolmogorov–Smirnov statistic, aggregated per duo and compared with paired *t*-tests.

*Software.* Analyses were implemented in Python using NumPy and SciPy (Virtanen et al., 2020), pandas, scikit-learn (Pedregosa et al., 2011), and statsmodels (Seabold and Perktold, 2010).

## Data and Code Availability

The Weimar Jazz Database is available at https://jazzomat.hfm-weimar.de. The

Improvising Duos corpus is available via https://doi.org/10.1371/journal.pone.0130070. The symbolic transcriptions, model outputs, and analysis code supporting this study are publicly available on the Open Science Framework (OSF) at osf.io/b9d3a.

## Acknowledgements

This research was supported by the JSPS KAKENHI (26K23870, 26H01202, 25K24741, 25K03206), Japan. The funding sources had no role in the decision to publish or prepare the manuscript.

## Author Contributions

T.D. conceived the project and designed the research. T.D. and Y.K. developed the theoretical framework. T.D. performed the analyses, interpreted the results, and supervised the study. R.K. and R.T. annotated the corpus data and contributed to the analytical design. Y.A. contributed to computational modeling and methodological development. T.D. developed the analytical pipeline, wrote the original draft, and prepared the figures. All authors contributed to the interpretation of the results and critically revised the manuscript. All authors approved the final manuscript. T.D. finalized the manuscript.

## Competing Interests

The authors declare no competing financial interests.

# Figure 1

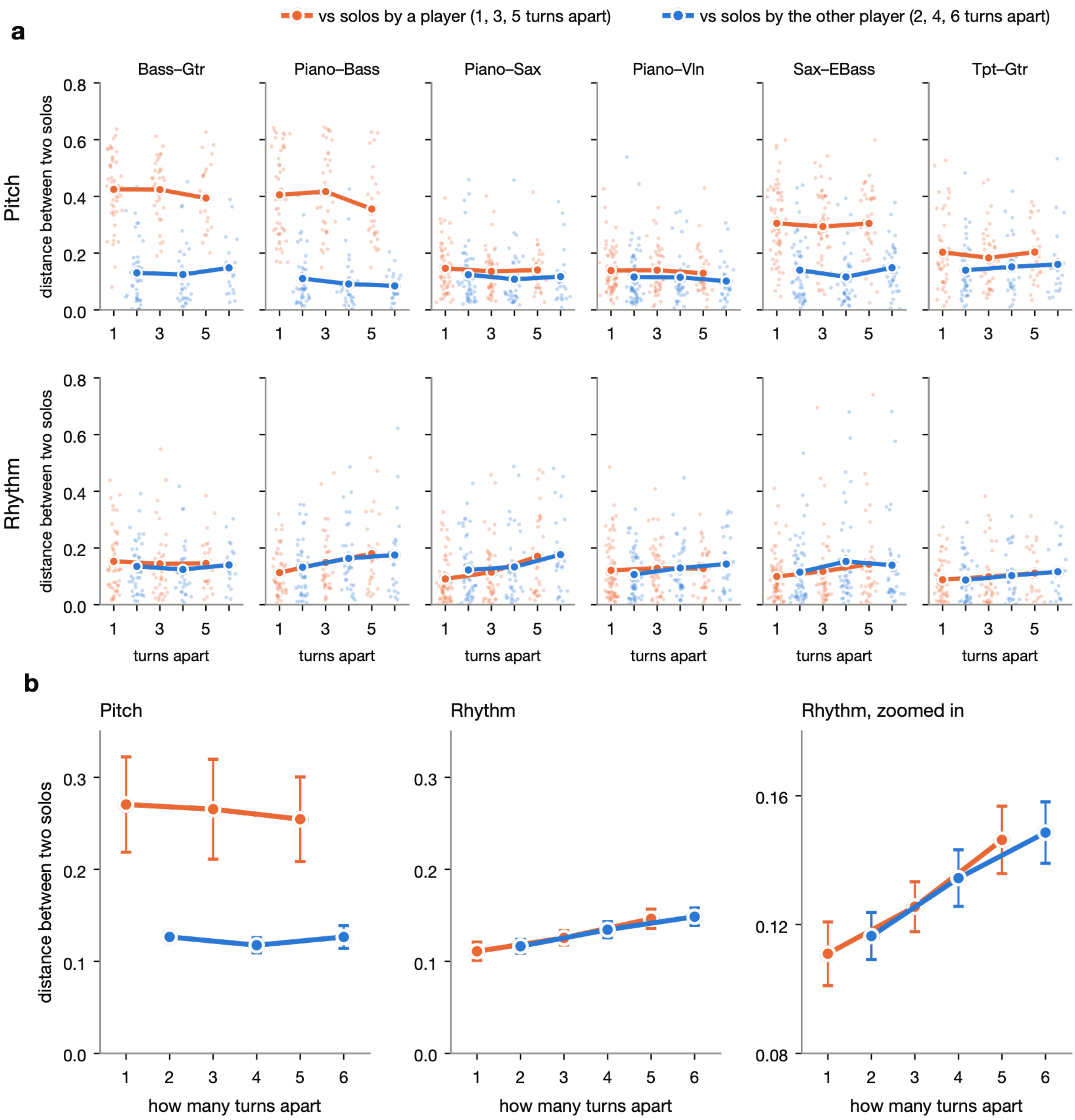

a
vs solos by a player (1, 3, 5 turns apart)
vs solos by the other player (2, 4, 6 turns apart)
Bass–Gtr
Piano–Bass
Piano–Sax
Piano–Vln
Sax–EBass
Tpt–Gtr
Pitch
Rhythm
distance between two solos
turns apart
b
Pitch
Rhythm
Rhythm, zoomed in
how many turns apart

**Figure 1. Melodic separation persists while the shared rhythmic state drifts.** (**a**) Each of the six duos, measured separately. Each small dot is one pair of solos from the same take; the distance between two solos is the Jensen–Shannon divergence between their note-level distributions of model-derived uncertainty. Orange, comparisons with the partner's solos (1, 3, or 5 turns apart); blue, comparisons with the performer's own previous solos (2, 4, or 6 turns apart); thick lines are the averages. Only takes long enough to supply every separation from 1 to 6 turns are included (41 of 46)[redacted], so that every point is computed from the same set of takes. In pitch (top), orange lies above blue in all six duos and the gap does not close; in rhythm (bottom), the two lines overlap and, in most duoes, rise together. (**b**) Group-level summary: thick lines, duo means; error bars, s.e.m. across duos. Left, pitch: a wide gap between partner and self comparisons that never closes. Middle, rhythm on the same scale: no gap—the two lines overlap. Right, rhythm magnified: both lines climb in parallel, indicating that each performer moves away from their own past at the same rate as from their partner; the shared state itself drifts.

# Figure 2

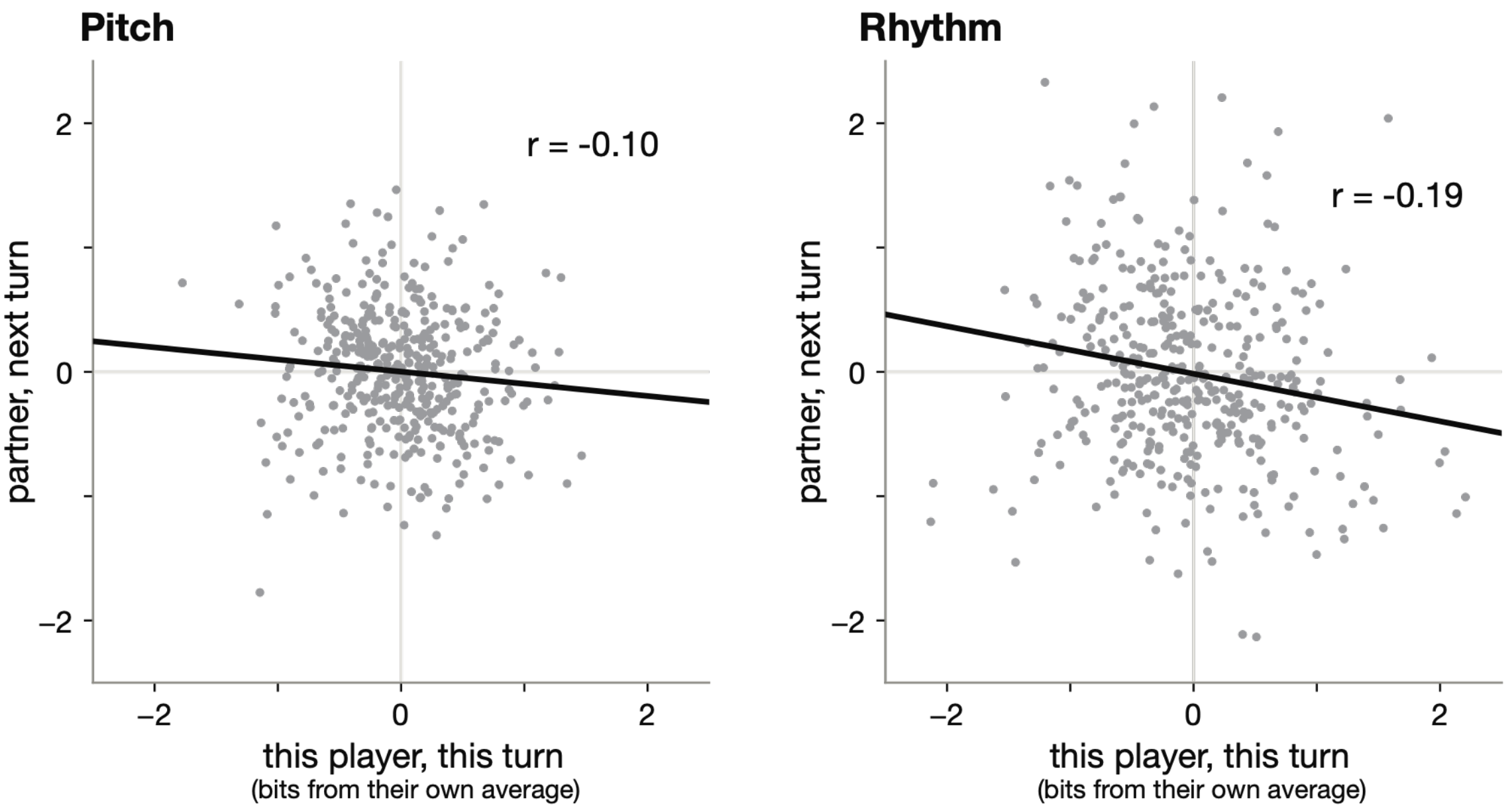

Pitch
r = -0.10
partner, next turn
this player, this turn
(bits from their own average)
2
0
−2
Rhythm
r = -0.19
partner, next turn
this player, this turn
(bits from their own average)

**Figure 2. At every hand-off, one performer pushes and the other pulls back**. Each dot is one hand-off between consecutive solos (*n* = 415 across the corpus). The x axis shows how far above or below their own average (in bits of surprise, after removal of take-level linear trends) the outgoing performer was on that turn; the y axis shows the same for the incoming partner on the very next turn. Downward-sloping clouds indicate complementarity: excess surprise by one performer is answered by restraint from the other. Both of the pitch and rhythm effects exceeded two null models that preserve each performer's own sequence while breaking its alignment with the partner (see Results).